\documentclass[aps,prl,twocolumn,groupedaddress]{revtex4}
\usepackage{graphicx}%
\begin{document}


\title{Evidence against the polarization rotation model of piezoelectric 
perovskites at the morphotropic phase boundary}


\author{J. Frantti}
\email[]{johannes.frantti@hut.fi}
\author{Y. Fujioka}
\author{R. M. Nieminen}
\affiliation{COMP/Department of Applied Physics, Helsinki University of Technology, FI-02015-HUT, Finland}


\date{\today}

\begin{abstract}
The origin of the very large piezoelectric response observed in
the vicinity of the morphotropic phase boundary (MPB) in perovskite 
lead zirconate titanate and related systems has been under intensive 
studies. 
Polarization rotation ideas are frequently invoked to explain 
the piezoelectric properties. It was recently reported
that lead titanate undergoes a phase transformation sequence
$P4mm\rightarrow Pm\rightarrow Cm\rightarrow R\bar{3}c$
at 10 K as a function of hydrostatic pressure [M. Ahart et al.
Nature Letters. \textbf{451}, 545 (2008)]. We demonstrate
that this interpretation is not correct by (i) simulating the
reported diffraction patterns, and (ii) by density-functional
theory  computations which show that neither the $Pm$, $Cm$ nor
$Pmm2$ phase is stable in the studied pressure range, and
further show that octahedral tilting is the key stabilization
mechanism under high pressure. Notes on a more general ground 
are given to demonstrate that a continuous
phase transition between rhombohedral and tetragonal phases via
intermediate monoclinic phase is not possible. Thus, two-phase
co-existence in the vicinity of the phase transition region is
probable and has an important role for electromechanical
properties.
\end{abstract}


\maketitle

The polarization rotation (PR) model \cite{Fu,Guo} has been proposed to 
explain the large electromechanical coupling coefficients observed in 
ferroelectric perovskites in the vicinity of the morphotropic phase 
boundary (MPB). The MPB region separates tetragonal and rhombohedral 
phases, which do not have a group-subgroup relationship and thus no 
continuous transition between the phases is possible. The most intensively 
studied systems are solid solutions, prime examples being lead zirconate 
titanate, Pb(Zr$_x$Ti$_{1-x}$)O$_3$, (PZT) and 
$x$Pb(Mg$_{1/3}$Nb$_{2/3}$)O$_3$-($1-x$)PbTiO$_3$ (PMN-PT). The essential 
feature of the PR model is the insertion of one (or more) 
low-symmetry phase(s) to continuously (via group-subgroup chains) connect the 
tetragonal and rhombohedral phases separated by the MPB in order to 
continuously rotate the polarization vector by an electric 
field or pressure between the pseudo-cubic [001] and [111] directions along 
the $(1\bar{1}0)$ plane. This rotation path was predicted to be accompanied 
by a large electromechanical response \cite{Cohen}. There are, however, 
several ambiguities related to the PR model (see, e.g., Ref. \onlinecite{Kisi}) 
and experimental studies interpreted in terms of this idea. As an example, 
the pressure induced phase transitions of lead titanate (PbTiO$_3$, PT) are 
considered below. Hydrostatic pressure induces similar structural changes as 
are observed to occur due to the substitution of Ti by a larger cation, such 
as Zr, causing so called ``chemical pressure''.

At high temperatures PT undergoes a phase transition between the $P4mm$
and $Pm\bar{3}m$ phases \cite{Sani}. At room temperature PT
transforms to a cubic phase through a second-order transition at
12.1 GPa\cite{Sanjurjo}, whereas it was predicted through 
density-functional theory (DFT) computations that a phase 
transition between
$P4mm$ and $R3c$ phases occurs at 9 GPa at 0 K \cite{FranttiJPCB}.
Notably the latter phase transition is similar to the phase
transition observed in PZT as a function
of Zr composition. In simplest terms, one expects to have three
different phase boundaries in the pressure-temperature plane of PT,
separating the $P4mm$ and $Pm\bar{3}m$, $P4mm$ and $R3c$ and $R3c$
and $Pm\bar{3}m$ phases. A very different interpretation was
recently given in Ref. \onlinecite{Ahart}, according to which the
phase transition from the $P4mm$ to $R\bar{3}c$ phase would occur
via monoclinic phases, which was further claimed to give support to
the PR model. We demonstrate that (i) the single
phase model is incorrect in the vicinity of the phase transition,
(ii) the monoclinic distortions reported earlier are not stable,
(iii) summarize the arguments which show that the phase transition
must be of first order and (iv) outline the method for determining
the piezoelectric properties in the vicinity of the phase boundary.

\paragraph{Computational methods.}
The DFT code ABINIT \cite{Gonze1,Gonze2} was used to compute the 
total energies and phonon frequencies and eigenvectors \cite{Gonze3} 
at different pressures. The computations were carried out within the 
local-density approximation and a plane wave basis. Norm-conserving 
pseudopotentials were generated using the OPIUM package \cite{Rappe}. A
more detailed description of the computational approach is available in Ref.
\onlinecite{FranttiJPCB}. For the simulation of the X-ray diffraction patterns
the Powder Cell program was used \cite{Krans}. The lattice parameters were
adapted from Ref. \onlinecite{Ahart}. The asymmetric unit was not given in
Ref. \onlinecite{Ahart}, and thus the atomic positions were estimated using the
values found from the DFT computations, which are close to the values estimated
from our high-pressure neutron powder diffraction experiments at few GPa
pressures \cite{FranttiLosAlamos}.

\begin{figure*}
\includegraphics[width=10.4cm]{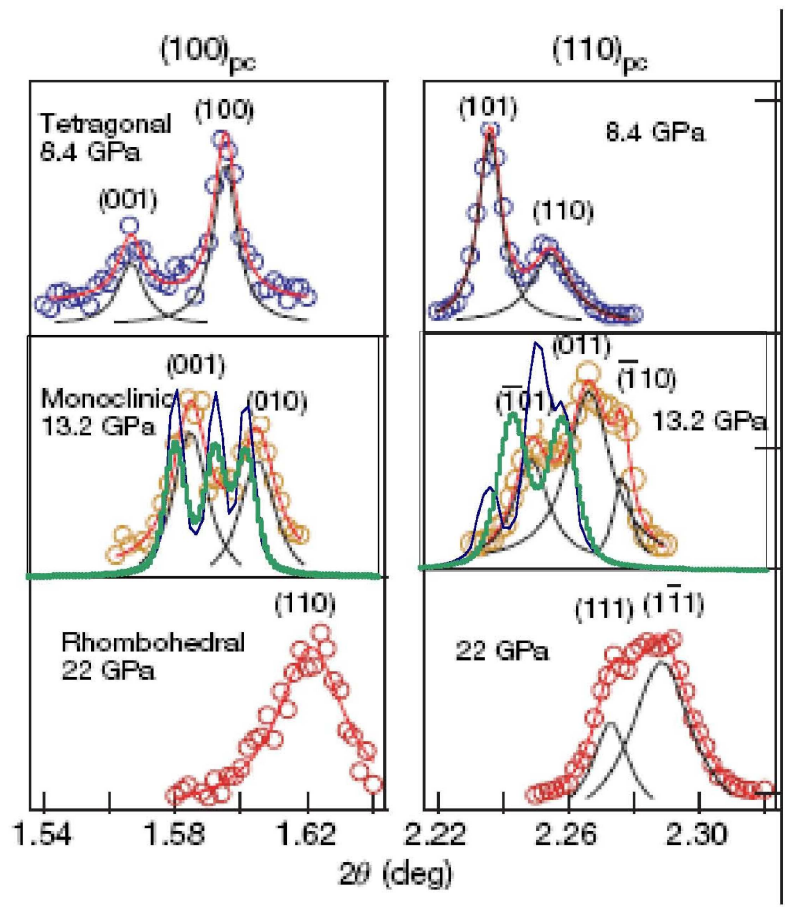}
\caption{\label{Simulations}X-ray diffraction data collected on PT
at 10 K. The figure is adapted from Ref. \onlinecite{Ahart}. The green
and blue lines (middle panels) were added by us. The green line shows
the simulated $Pm$ pattern using the lattice parameters given in
Ref. \onlinecite{Ahart}. The model where the $a$ and $b$ axes are
switched (blue line) does not improve the fit \cite{Note1}. Neither
of the one-phase $Pm$ structure model fits the peak (black lines)
positions and intensities (e.g., the reflection labelled as
$(\bar{1}10)$ is not modelled, and cannot be explained by preferred
orientation).}
\end{figure*}

\paragraph{Notes on the X-ray diffraction and Raman scattering analysis.}
According to Ref. \onlinecite{Ahart}, PT undergoes a phase transformation
sequence $P4mm\rightarrow Pm \rightarrow Cm \rightarrow R\bar{3}c$
at 10 K as a function of hydrostatic pressure. We show that the
X-ray diffraction (XRD) pattern collected at 13.2 GPa \cite{Ahart}
is not consistent with the reported $Pm$ symmetry by simulating the
corresponding pattern. Fig. \ref{Simulations} shows that the
reflection positions and intensities significantly deviate from the
experimental ones and also from the fits (shown by black continuous
lines). It is worth to note that in the case of PT the pseudo-cubic
110 reflections have the strongest XRD intensities. The 13.2 GPa XRD
pattern shown in Fig. \ref{Simulations} more likely corresponds to a
two-phase diffraction pattern. This is seen by studying the
intensities of the 100 and 001 reflections: for tetragonal and
pseudo-tetragonal structures the intensity ratio should roughly be
2:1 (as is seen from the diffraction pattern collected at 8.4 GPa,
Fig. \ref{Simulations}), whereas it is roughly 0.9:1 for the 13.2
GPa data.

It was stated that the Raman scattering data reflect the monoclinic
$M_C$ ($Pm$ phase) to monoclinic $M_A$ ($Cm$) and the monoclinic
$M_A$ to rhombohedral phase transitions \cite{Ahart}. We find this
assignment questionable, since the phonon symmetries, central for
the phase transition studies, were not addressed. For example, the
$B_1$-symmetry normal mode in the $P4mm$ phase breaks the fourfold
symmetry \cite{FranttiJJAP}, whereas the $A_1$ symmetry modes
preserve it. The spectral features below 100 cm$^{-1}$ include
several peaks from the $A_1$ symmetry modes alone, due to the strong
anharmonicity of the $A_1$(1TO) mode \cite{Foster1,Foster2}, in
addition to the $E$-symmetry modes and Rayleigh scattering (which
dominates the region close to the laser line, as was noted in Ref.
\onlinecite{Sanjurjo}). It was rather recently that the $A_1$(1TO)
mode was identified in PT\cite{Foster1,Foster2}: many earlier
assignments dismissed this mode since the line shape was very
asymmetric and turned out to be consisted of many subpeaks. In
practice this means that, in the vicinity of the phase transition, it
is hard to identify the number of modes at the low-frequency region,
not to mention the difficulty of identifying their symmetries from
the spectra collected without proper polarization measurements.
This, in turn, prevents space group assignments.

\paragraph{DFT studies.}
DFT computations predict that PT undergoes a phase transition from
the $P4mm$ phase to the $R3c$ phase at around 9 GPa
\cite{FranttiJPCB}. In contrast, a phase transition sequence
$P4mm\rightarrow Cm \rightarrow R3m \rightarrow Pm\bar{3}m$ (phase
transitions at 10, 12 and 22 GPa, respectively) was found in Ref.
\onlinecite{Wu}. The high-pressure end of this transition was more
recently modified to form the sequence 
$R3m \rightarrow R3c \rightarrow R\bar{3}c
\rightarrow R3c$ with phase transitions occurring at 18, 20 and 60
GPa, respectively \cite{Ahart}. In addition to the phases listed in
Ref. \onlinecite{FranttiJPCB}, we carried out similar computations
for the $Pm$ and $Pmm2$ phases. For consistency, phonon frequencies
of the $R3c$ phase were computed at 9, 10 and 15 GPa pressures at the
Brillouin zone center and boundary points. 

\begin{figure}
\includegraphics[width=7.4cm]{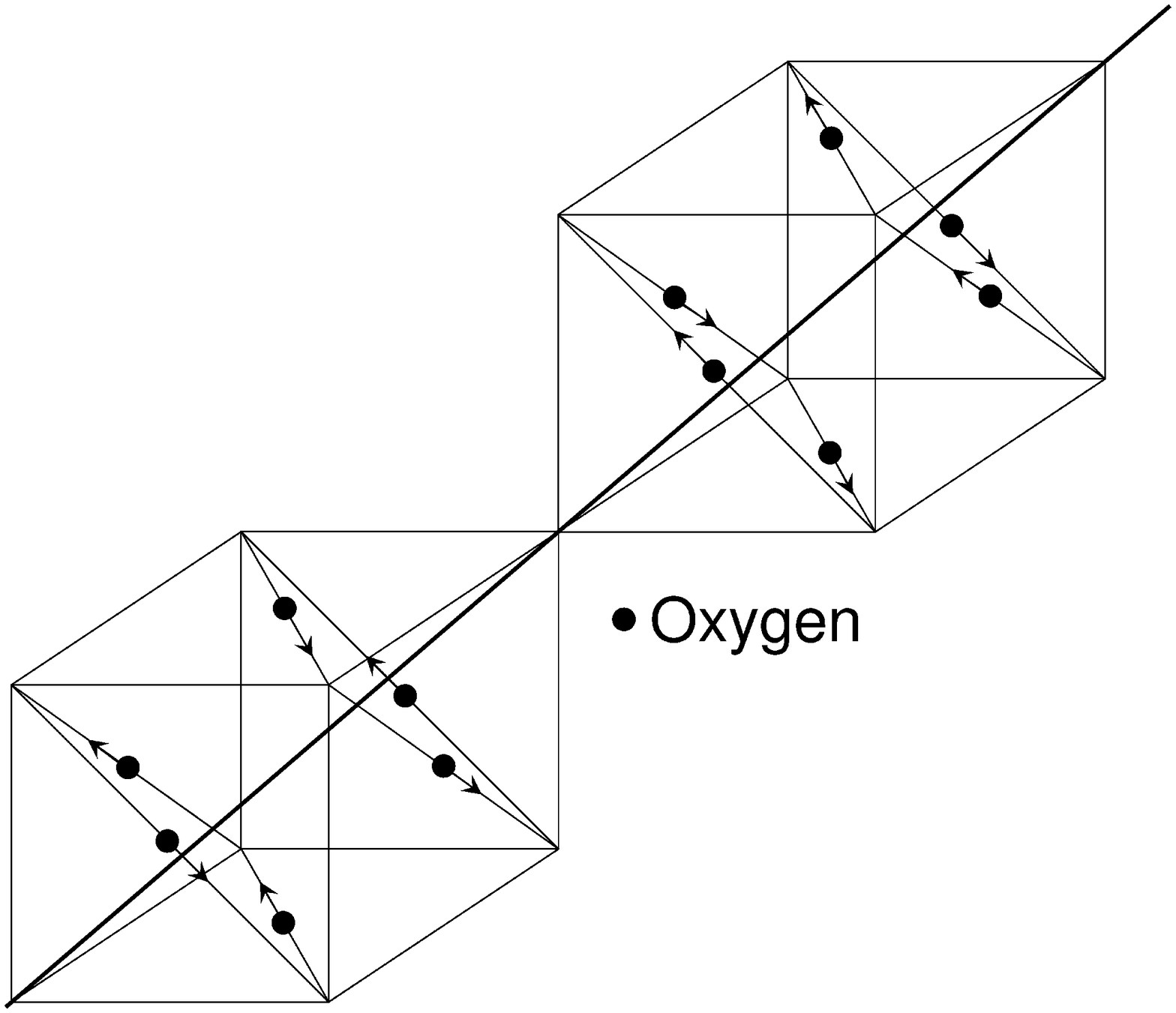}
\caption{\label{Octahedra_Rotation}. The unstable normal mode of the
$R3m$ phase at the $L = (\frac{\pi}{a}\frac{\pi}{a}\frac{\pi}{a})$
point involves only oxygen ions. Two rhombohedral unit cells are
shown: it is seen that the two octahedra are tilted about the
threefold axes clockwise and anticlockwise. The condensation of this
mode corresponds to the phase transition $R3m\rightarrow R3c$. All
the modes had positive frequencies when the $R3c$ phase was used. 
The bold line is the threefold rotation axis.}
\end{figure}

The main outcomes of our
present and earlier computations are: (i) the $R3m$ phase is not
stable (octahedral tilting makes $R3c$ phase favorable above 9 GPa),
(ii) above 9 GPa tetragonal ($P4mm$ and $I4cm$), orthorhombic
($Cmm2$ and $Pmm2$) and monoclinic ($Pm$ and $Cm$) phases were
revealed to be unstable by the Brillouin zone boundary modes and
higher enthalpy values, (iii) no support for an intermediate phase
was found, and (iv) no phonon instabilities were observed in the
$R3c$ phase. In contrast, one of the Brillouin zone corner point
$L= (\frac{\pi}{a}\frac{\pi}{a}\frac{\pi}{a})$ modes of the $R3m$
phase was unstable at 9 GPa pressure. The mode involved only oxygen
displacements (this was the only mode which was found to be
unstable: all modes at the $(000)$, $(0 0 \frac{\pi}{a})$ and
$(\frac{\pi}{a}\frac{\pi}{a} 0)$ symmetry points were positive). The
mode is depicted in Fig. \ref{Octahedra_Rotation}. This corresponds
to the mode were the upper and lower octahedra are tilted clockwise
and anticlockwise about the threefold symmetry axis, thus again
demonstrating that the octahedral tilting stabilizes the $R3c$
phase. This is due to the fact that octahedral tilting allows a more
efficient compression \cite{Thomas,FranttiJPCB,FranttiPRB}. 

We note
that since the $R3m$ phase is not stable, it is somewhat hypotetical
to consider the instability of an unstable phase. A more rigorous
treatment, starting from the $P4mm$ phase, is given in Ref.
\onlinecite{FranttiJPCB}, with the same outcome. Thus, the energetically
favorable phase was obtained by allowing the crystal to relax
according to the normal mode displacements of the unstable modes
seen in the $P4mm$ phase. Thus the transition between $P4mm$ and
$R3c$ phases is characterized by two-phase co-existence, in an
analogous way to the phase transitions seen in PZT as a function of
composition. This is an important prediction as it in turn suggests
that the two-phase co-existence has a crucial role for the
piezoelectric properties near the phase transition pressures in PT,
in a similar way as was demonstrated in Ref. \onlinecite{Li} for PZT
in the vicinity of the MPB.

\paragraph{Symmetry considerations.}
Group-theoretical analysis indicates that, although the phase
transition between monoclinic and tetragonal phases can be continuous, 
the transition between rhombohedral and monoclinic phases must
be of first order \cite{Sergienko}. Thus, even if one would have a
monoclinic phase, it would not make the transformation path continuous.
First-order transitions are often characterized by the two-phase co-existence,
one phase being metastable over a finite temperature or pressure range.
This is consistent with the experimentally known features of PZT according
to which there is two-phase co-existence \cite{Cox,FranttiPRB,FranttiJPCM}.
Neutron and X-ray powder diffraction studies revealed that the polarization
vector in the monoclinic $Cm$ phase is very close to the pseudo-cubic
[001] direction, and hardly rotates from that direction
\cite{FranttiPRB,FranttiJPCM}, in contrast to what one anticipates from
the PR model. Thus the polarization vector changes
discontinuously when the transition from the pseudo-tetragonal monoclinic
to the rhombohedral phase occurs. As Li et al. noted, ``the availability of
multiple phases at the MPB makes it possible for the polarization to thread
through the ceramic''\cite{Li}.

\paragraph{How to model the piezoelectric response?}
The piezoelectric response can be divided to extrinsic and intrinsic
contributions. The latter is due to the changes in electron densities
as a response to an applied field or stress and can be computed through
standard density-functional theory methods. The extrinsic part is
significantly more challenging, as it involves domain wall motions and
changes in the phase fractions in the vicinity of the phase-boundary
(e.g., between tetragonal and rhombohedral phases).
In the case of poled ceramics one first computes the necessary angular
averages of the piezoelectric constants and takes their dependence on
temperature, composition or stress into account. This dependence is notable in
the vicinity of the phase transition. For intrinsic contribution such
a computation is rather straightforward. However, the description of
domain wall motion due to an applied electric field or stress for
different composition or at different temperatures is nontrivial task.

In conclusion, evidence against the applicability of the polarization 
rotation model to perovskites is strong. Instead, the currently known best 
piezoelectric perovskites posses a so-called morphotropic phase boundary at 
which a first-order phase transition between rhombohedral and tetragonal 
(or pseudo-tetragonal) phases takes place. For the electromechanical properties 
it is important to note that this transition exhibits two-phase co-existence. 
Structural factors responsible for the stabilization of the rhombohedral phase, 
either at large hydrostatic pressures or large chemical pressures (as occurs 
in Pb(Zr$_x$Ti$_{1-x}$)O$_3$ with increasing $x$) were addressed.

\begin{acknowledgments}
Y. F. is grateful for the Finnish Cultural Foundation and Tekniikan
Edist\"amis\"a\"ati\"o foundation for financial support. This project
was supported by the Academy of Finland (Project Nos 207071
and 207501 and the Center of Excellence Program 2006-2011). 
Finnish IT Center for Science (CSC) is acknowledged for
providing computing resources.
\end{acknowledgments}
\bibliography{Frantti}

\end{document}